# 4MOST Low Resolution Spectrographs Characterization in Chile

Florence Laurent*[a], Didier Boudon[a], Eric Daguisé[a], Rémi Giroud[a], Aurélien Jarno[a], Alexandre Jeanneau[a], Andreas Kelz[b], Jens-Kristian Krogager[a], Victor Mauger-Vauglin[a], Jean-Emmanuel Migniau[a], Matthew Lehnert[a], Flora Paganelli[a], Arlette Pécontal[a], Emmanuel Pécontal[a], Alban Remillieux[a], Johan Richard[a], Allar Saviauk[b, c]

[a] Universite Claude Bernard Lyon 1, CRAL UMR5574, ENS de Lyon, CNRS, Villeurbanne, F-69622, France

[b] Leibniz-Institut für Astrophysik Potsdam (AIP), An der Sternwarte 16, D-14482 Potsdam, Germany

[c] European Southern Observatory (ESO), Karl-Schwarzschild-Str. 2, Garching bei Munchen, Germany

**ABSTRACT**

4MOST, the 4m Multi Object Spectroscopic Telescope, is the optical, fibre-fed, MOS facility for the VISTA telescope at ESO's Paranal Observatory in Chile. Its main science drivers are in the fields of galactic archeology, high-energy physics, galaxy evolution and cosmology. The 4MOST consortium consists of several institutes in Europe and Australia under leadership of the Leibniz-Institut für Astrophysik Potsdam (AIP).

This paper focuses on the successful testing, installation and technical commissioning of the Low Resolution Spectrographs (LRS-A and B) for the 4MOST instrument at ESO's Paranal Observatory, Chile. This work was completed on October 18. 2025. Details on the assembly, integration, and performance of both 4MOST spectrographs from their arrival in the integration hall through to the telescope installation are provided. Attention is given to the optimization of procedures implemented to enhance performance and meet the expected top-level requirements.

The 4MOST LRS features 2436 fibres split into two low-resolution spectrographs LRS-A and LRS-B (1624 fibres, three arms, 370-950 nm, R > 4000) and one high-resolution spectrograph (812 fibres, three arms, ~44-69 nm coverage each, R > 18000). The fibre positioner covers a hexagonal field of view of ~4.1 deg². The fibers are 85 µm core with an output beam at f/3. The Centre de Recherche Astrophysique de Lyon (CRAL) had the full responsibility for the two low-resolution spectrographs. Each of them is composed of an off-axis Schmidt collimator that produces a 200 mm beam, which is split into three spectral arms by dichroics and directed to F/1.73 cameras with standard 6k x 6k 15µm pixel CCD detectors.

## 1 INTRODUCTION

The 4MOST Low Resolution Spectrographs (LRS) successful testing, installation and technical commissioning at ESO's Paranal Observatory, Chile, was completed on October 18, 2025.

The local acceptance reviews for both 4MOST LRS was conducted at CRAL and were successfully completed in December 2021 and October 2022, respectively. In 2022, two SPIE papers - [3][5],[6] - detailed the processes carried out at CRAL, from the integration and alignment of the sub-assemblies to the test procedures used to demonstrate the spectrographs' compliance with requirements. In 2024, another SPIE papers - [7] , [8] - focused on the re-alignment and verification of the spectrographs in Europe, at Potsdam.

Following the successful Provisional Acceptance in Europe (PAE), in May 2025, by the 4MOST's Project Office and ESO for the whole 4MOST Consortium, the two low-resolution spectrographs units were partially disassembled and shipped to Chile. During the summer of 2025, the CRAL team completed their realignment in the new Integration Hall at Paranal, after which they were installed on the VISTA telescope.

*florence.laurent@univ-lyon1.fr; phone +33 4 78 86 85 33; http://cral.osu-lyon.fr/

The paper outlines the 4MOST LRS successful testing, installation and technical commissioning at ESO's Paranal Observatory, Chile, providing details on the assembly, integration, and performance of both 4MOST LRS from their arrival in the integration hall at Paranal through the installation at the VISTA telescope. Attention is given to the optimization of procedures implemented to enhance performance and meet the expected top-level requirements.
Following are sections dedicated to the 4MOST design and requirements (section 2), the schematic strategy followed for the spectrographs' characterization at Paranal (section 3), details on dismount in Potsdam for transport to Chile (section 4), the phase of the alignment and integration and verification at Paranal (section 5 and 6), and the phase related to the installation on VISTA and final technical commissioning (section 7).

## 2 LRS DESIGN AND KEY REQUIREMENTS

The LRS is composed of two symmetrical unit spectrographs including the same optical components. Each spectrograph should meet the key requirements listed hereafter. Each spectrograph is fed by an entrance slit of 812 science fibres. After collimation, the optical beam is separated in three spectral bandwidths, dispersed and reimaged in three different channels with F/1.73 aperture. The separation in three channels provides the minimal resolution of 4000 (Figure 1).
The Blue arm goes from 370 to 554 nm, the Green from 524 to 721 and the Red from 691 to 950 nm.
The LRS is composed of the 7 sub-assemblies:

- An Entrance Slit Unit (ESU) composed of 812 science fibres following focal plane curvature radius [3]. A Slit Field Lens is connected, using index-matching gel, onto the entrance slit. Note that the LRS Manufacturing, Alignment, Integration, Test (MAIT), Assembly Integration and Testing (AIT), and Assembly Integration and Verification (AIV) phases were performed with a Test Slit Unit (TSU), hosting a partially populated slit.
- A spherical mirror and correctors act as a Schmidt collimator collecting the F/3 light from the fibres and generates a collimated beam of 200mm.
- A Dichroic: Dichroic beam splitters in front of the Schmidt corrector separate the three channels. There are two dichroics (Red and Blue).
- In addition, each channel includes a corrector, a Volume Phase Holographic Grating (VPHG) for dispersion, a dioptric camera and a Field Lens Window. The spectral overlap of the channels is determined by the wavelength cutoff and here assumed to be 30nm. A 6k x 6k 15µm/pixel CCD detector records the spectra.

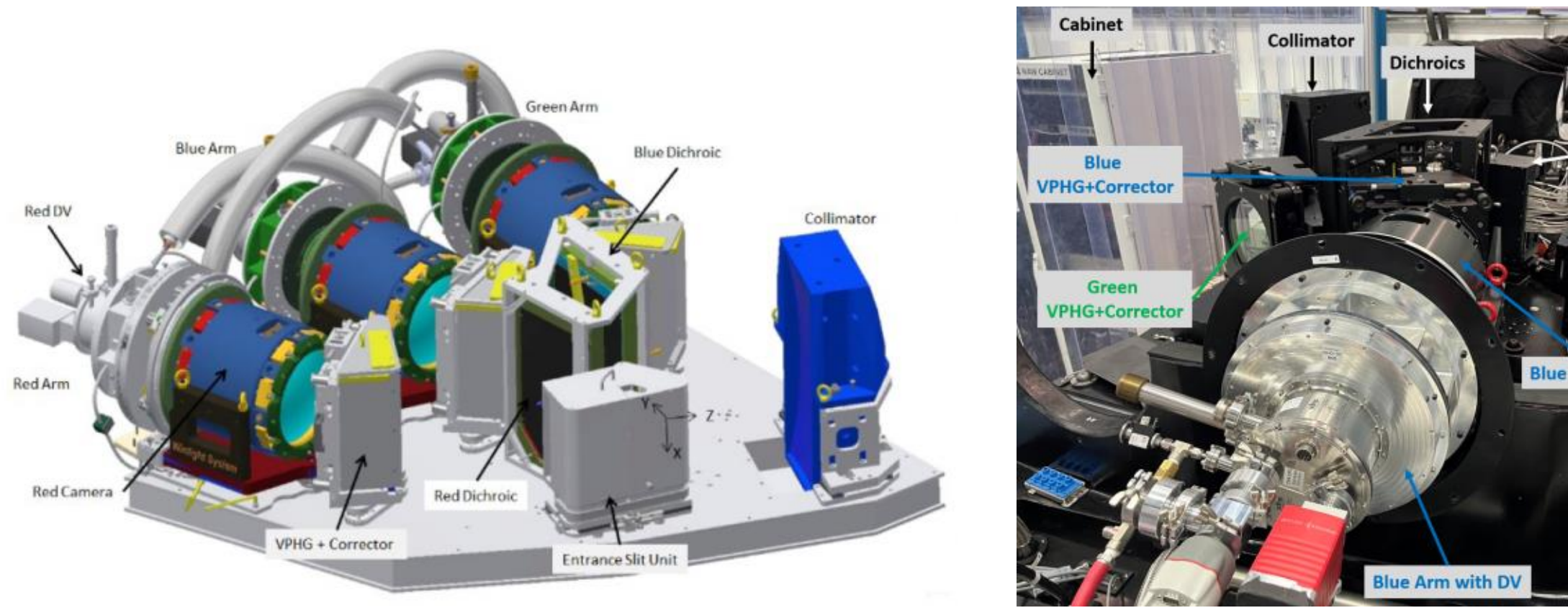

Figure 1: Left: LRS mechanical model. Right: LRS-A and -B being integrated in Paranal Integration Hall.

The LRS key design requirements are derived from the 4MOST top level requirements - [1], [2] - and are the following:

1. The LRS shall be able to accommodate a minimum of 812 science fibres and 10 calibration fibres. It shall accept f/3.0, with 85 microns core diameter fibre input. The LRS shall cover simultaneously the wavelength range from 400 nm to 885 nm (goal: from 370 nm to 950 nm). The LRS spectra shall be projected on 6k by 6k, 15 micron pixel detector(s).
2. The LRS spectral resolving power shall be $R \geq 10 \times \lambda$/nm for 400 nm $\leq \lambda <$ 500 nm, and $R \geq 5000$ for 500 nm $\leq \lambda \leq$ 885 nm.
3. The LRS spectral sampling shall be $\geq$ 2.5 pixel (goal 3.0 pixels).
4. Cross-talk between spectra: At any given wavelength and at all locations of spectra, no spectrum shall contaminate its neighbouring spectra on the detector with more than 2% (goal 1%) of its own flux.

## 3 LRS DEPLOYMENT STRATEGY

The Assembly, Integration, Verification (AIV) procedure at Paranal was divided into two main phases: the first one occurred at the Paranal New Integration Hall (NIH), and the second at the VISTA telescope. The activities conducted through the deployment are outlined in the LRS AIV Flow Chart shown in Figure 2.

The first phase consisted in the remounting, alignment and health checks of both LRSs in using the Test Slit Unit (TSU). The activities were the same as those performed during the Alignment, Integration and Tests (AIT) at AIP [7]. The MAIT carried out at CRAL was designed to be the same as the one at AIP and Paranal. The goal was to use the same AIT tools, mounting, alignment and verification procedures implemented at CRAL to minimize integration activities at Paranal. At the end of AIV Phase 1, an LRS Installation Readiness Review gave the green light to transport the LRSs at the telescope.

The second phase consisted in the installation of the LRSs on the VISTA telescope. The installation of the ESU in LRS, the technical commissioning and first health checks using the sky and the Calibration System.

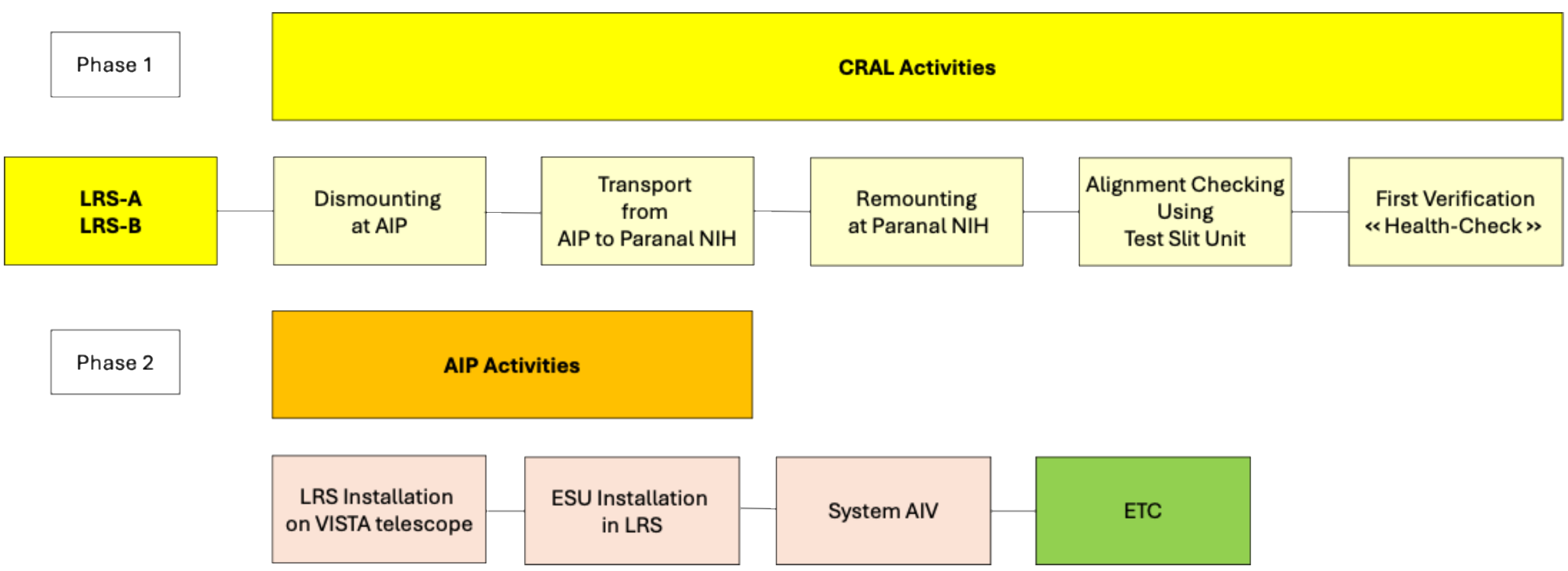

Figure 2: LRS AIV Flow Chart: Phase 1 - CRAL activities in the Paranal New Integration Hall (NIH); Phase 2 - AIP activities leading to the End of Technical Commissioning (ETC).

## 4 DISMOUNTING AT AIP AND TRANSPORT FROM POTSDAM TO PARANAL

### 4.1 Dismounting at AIP

The assembly was dismounted using Point/Line/Plane (PLP) interfaces at AIP. For LRS, the parts and assemblies have been delivered individually and packed in their individual transport boxes. The electronics, the detectors, and all associated AIT tools were packed by the MPIA and AIP teams. For the LRSs, a total of 18 crates - identical to those used for the CRAL/AIP transport - were used. The dismounting procedure remained the same as the one used during the MAIT phase. Some sample parcels are shown in Figure 3.

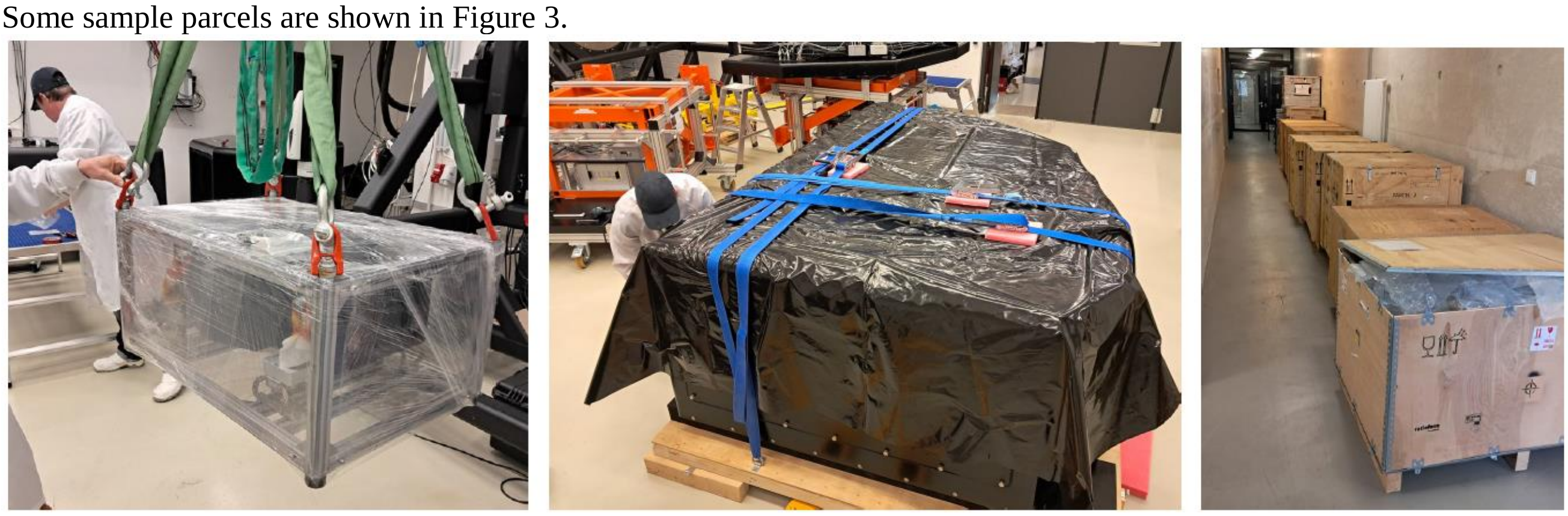

Figure 3: Left: Parcel#5 – Collimator. Middle: Parcel#15 – Main Structure. Right: Many parcels in the corridor.

### 4.2 Transport from AIP to Paranal NIH

All containers were equipped with appropriate sensors/indicators (temperature sensors, shocks recorder, tilt indicators) to detect any mishandling of the equipment during transportation. For relatively insensitive components, the standard transport wooden or aluminum crates were used. The very sensitive optics such as collimator, dichroic, dispersive assembly and Detector Vessel -DV were mounted on dedicated damping systems in plastic boxes. All crates were shipped by cargo plane from Frankfurt to Santiago de Chile and then transported by truck from Santiago to Paranal. Although tarps were normally expected when transporting such sensitive cargo, the trucks were not equipped with them. The unloading of the trucks was completed at Paranal NIH (Figure 4, left). There was a lot of rain during the trip to Paranal, and one of the collimator boxes got soaked. Fortunately, the collimator itself seemed to remain dry (Figure 4, right).

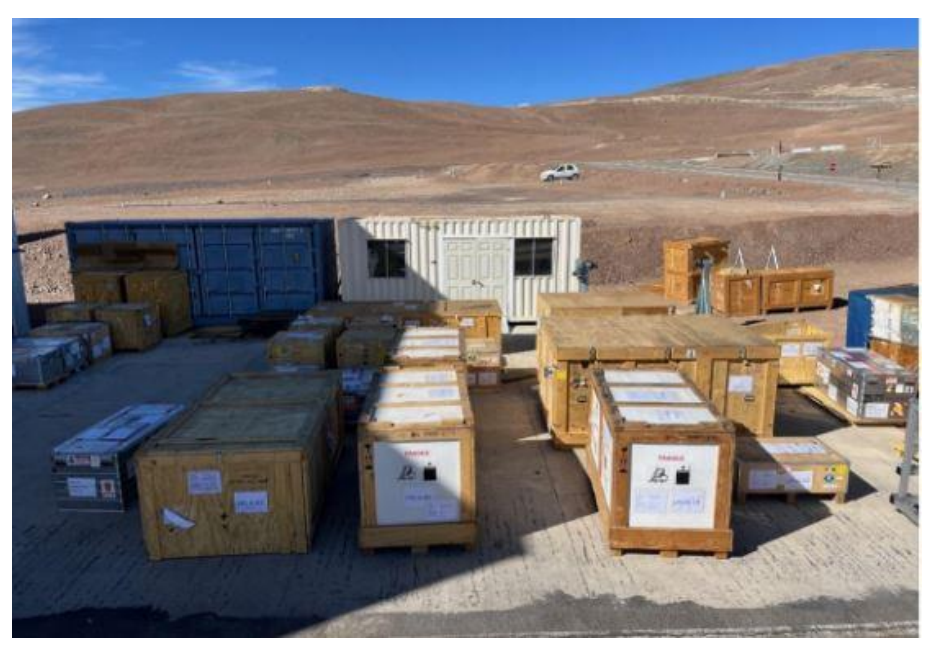

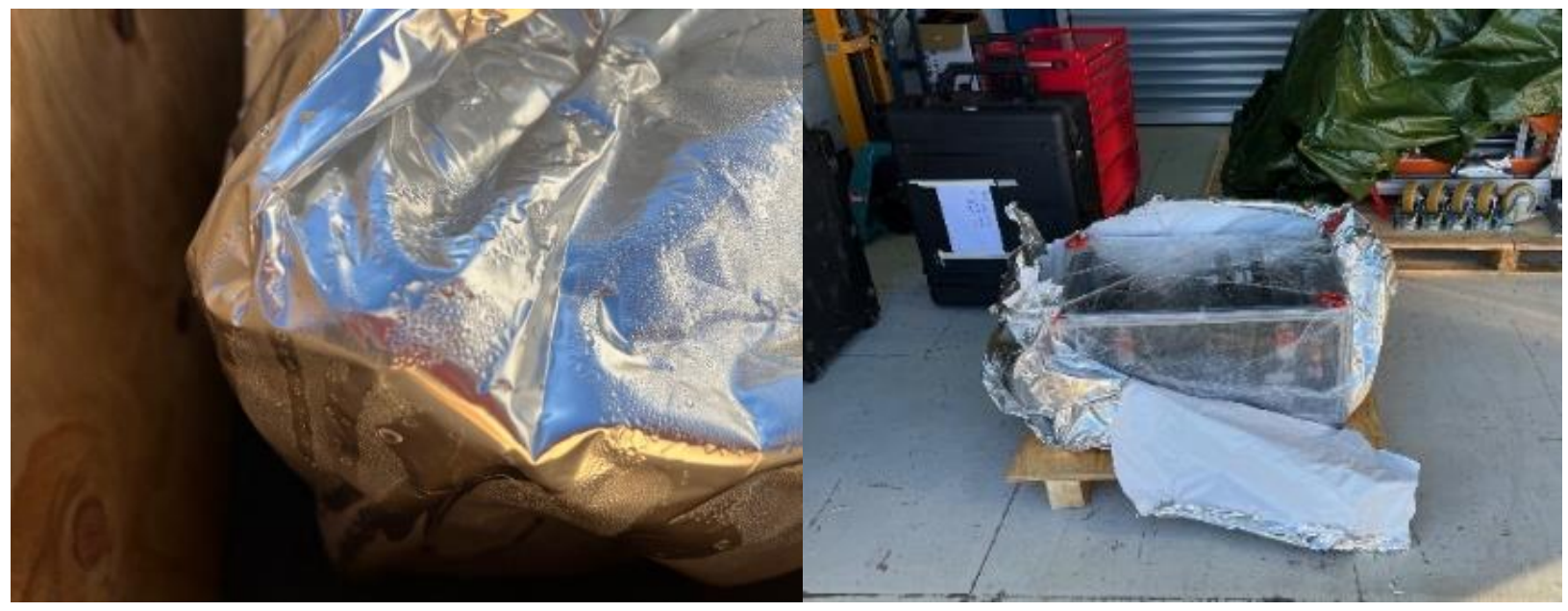

Figure 4: Left: LRS crates in front of the Paranal NIH. Right: Collimator crate

### 4.3 Remounting at Paranal NIH

The remounting at Paranal took place in the NIH (Figure 5). The clean tent and the frames have been placed in the same configuration as the telescope (spaced 260mm apart). The 2 frames have been anchored to the ground with 4 points, and optical bench A&B have been mounted onto their frame with a torque of 150 Nm each.

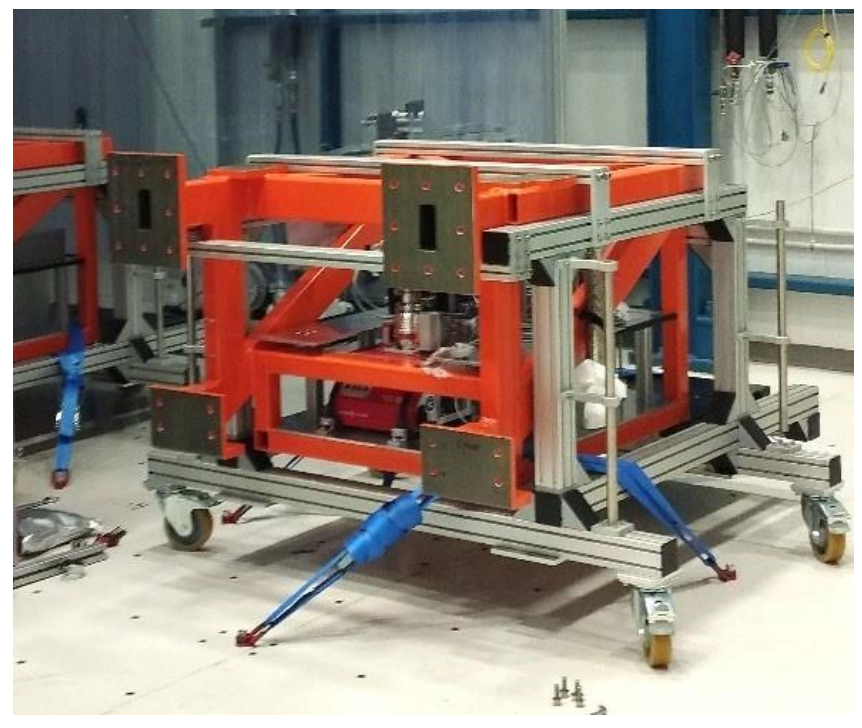

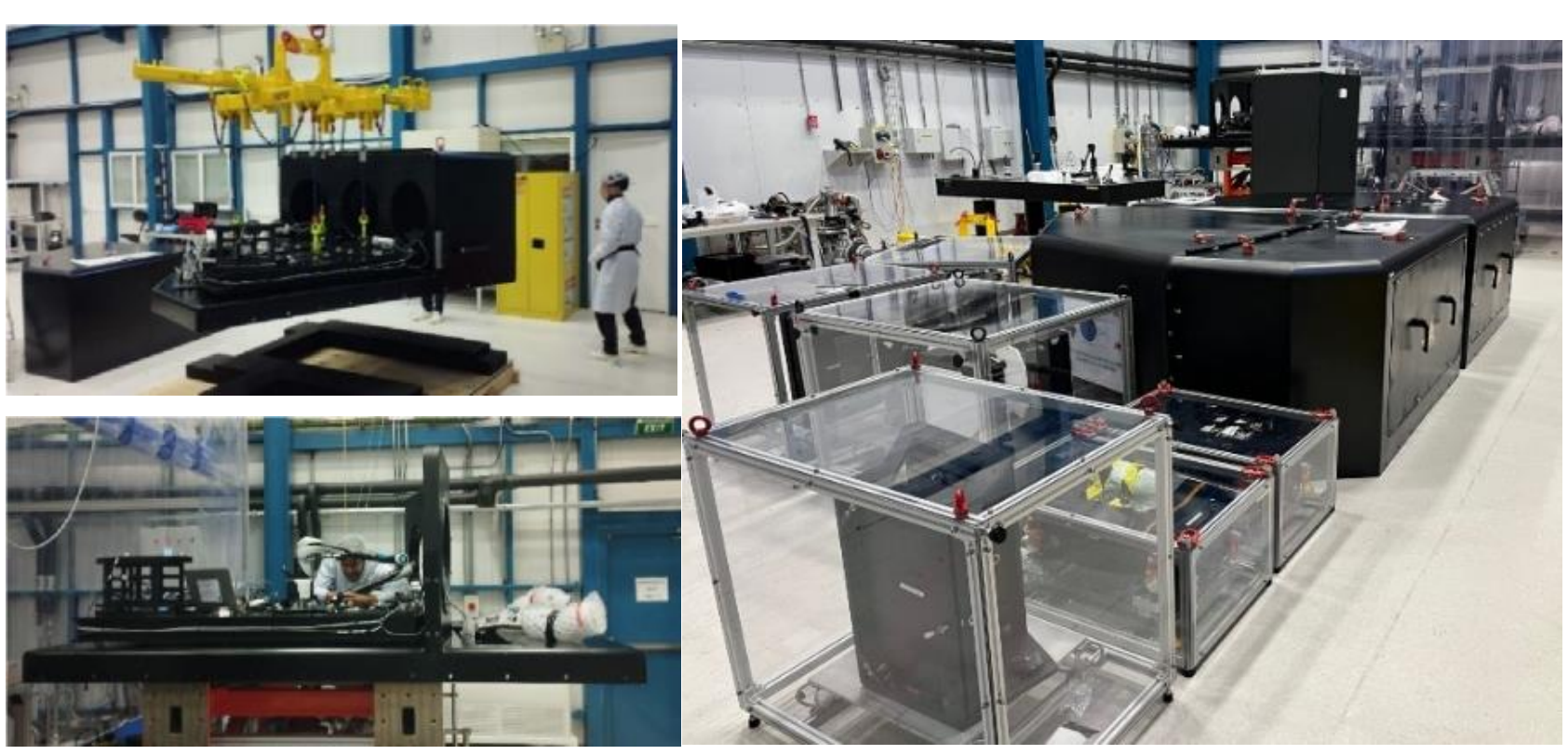

Figure 5: Left: The 2 frames has been anchored to the ground with 4 points. Middle: Mounted Optical bench with the Spreader and first measurement with the Hexagon arm. Right: Dichroics and Dispersive assemblies in the NIH

## 5 ALIGNMENT USING TEST SLIT UNIT

The papers [5] and [7] describe all the procedures implemented on the LRS MAIT at CRAL and on the LRS AIT at Potsdam; they outline the main MAIT tools used for LRS alignment, as well as the alignment procedure.

### 5.1 Common Path Alignment

After the installation of the LRS Frame, the Mounted Optical bench without optics and the cabinet, the Common Path Alignment has been performed. This procedure allows to align and verify the common path which includes Optical Axis Definition, Test Slit Unit, Collimator and Dichroics installation. This procedure has been repeated during the AIT at AIP and now in Chile. First, the measurement of the Point-Line-Plane (PLP) interfaces on the optical bench with Hexagon arm on both LRSs was measured. The x, y, z positions are like the one measured during AIT at AIP. Secondly, after their installations, a small correction on the collimator and both dichroics were performed in order to be centered on the alignment targets.

### 5.2 Arm & DV Alignment

The “Arm & DV” Alignment allows to align and verify the “arm” which includes Corrector, VPHG and Camera alignment. We start with the Blue arm because it is the least accessible. The procedure is then repeated for the Green and Red arms. To reach requirements, the adjustment in (z, Rx, Ry) of the Detector Vessel has been performed in using Test Slit Unit. The Test Slit Unit (TSU) hosts a partially populated slit; the slit box is thus composed of only 6 slitlets, which are themselves only sparsely populated with 11 fibers each [3]. In fact, during AIV at Chile, the (z, Rx, Ry) DV alignment was significantly improved. Now, these DV tip-tilt adjustments are based on focus curves and a compromise between spatial/spectral image quality and crosstalk performance, which allows us to determine the best focus plane. It was specifically for this reason that the re-alignment of the 6 DVs was redone at Paranal. Once the DV alignment was completed, the Top and DV Covers were installed and the Health checks for analysis was then started.

## 6 PERFORMANCE VERIFICATION

During the AIV at Chile, only the « Health-Check » has been performed. That allows to validate the re-alignment with the image quality, crosstalk and spectra positioning. The Interface verification was checked at Potsdam with the other subsystems by the system team [7]. The global tests for wavelength range, dichroic transition region, throughput, ghosts and straylight, spectrograph stability and temperature difference requirement are not reproduced at Chile because they have already validated at CRAL. The whole performance reached during the MAIT at CRAL are fully described in the paper [5].

### 6.1 Performance using arc lamps: Spectral resolving power and sampling

The LRS designed key requirements, as listed in Section 2, were checked at the Paranal NIH prior to the installation on VISTA. Analyses of the median final exposures were performed using the following procedure: 5 exposures each of the spectral arc lamps and fibre flat were taken for each LRS, bias subtracted, and the median of 5 exposures computed and analysed. The spectral lamp used for the exposures, as well as the light guides (Series 380 and 2000) between the lamp and the illumination unit were provided by AIP. A Hg(Ne) pencil lamp was used for arc exposures in the blue channel (Pencil calibration lamp from LOT-Oriel with 18mA power supply), while a Ne pencil lamp (Pencil style calibration lamp from LOT-Oriel with 10 mA power supply) was used for arc exposures in the green and red channels.

For the spectral resolving power and sampling analyses, the Line Spread Function (LSF) used a custom model, which convolved a fibre model with a PSF modelled as a circular 2D Gaussian parametrized by its FWHM. The FWHM found for the best fitted model gave the size of a resolution elements, which represents the spectral sampling in pixels. The spectral resolving power R is the ratio between the wavelength and the FWHM. The spectral resolution obtained for LRS-A and LRS-B at Paranal, and comparison with previous AIT phases, is shown in Figure 6 and Figure 7. There are two graphs side by side: one shows the spectral resolution and the other shows the spectral sampling. The x-axis represents the wavelength range and the y-axis represents the spectral resolution and spectral sampling respectively. The solid line represents the requirements and the dotted line is the goal. The figures clearly show the improvement in results since the MAIT in Lyon. Everything is now in line with the specifications.

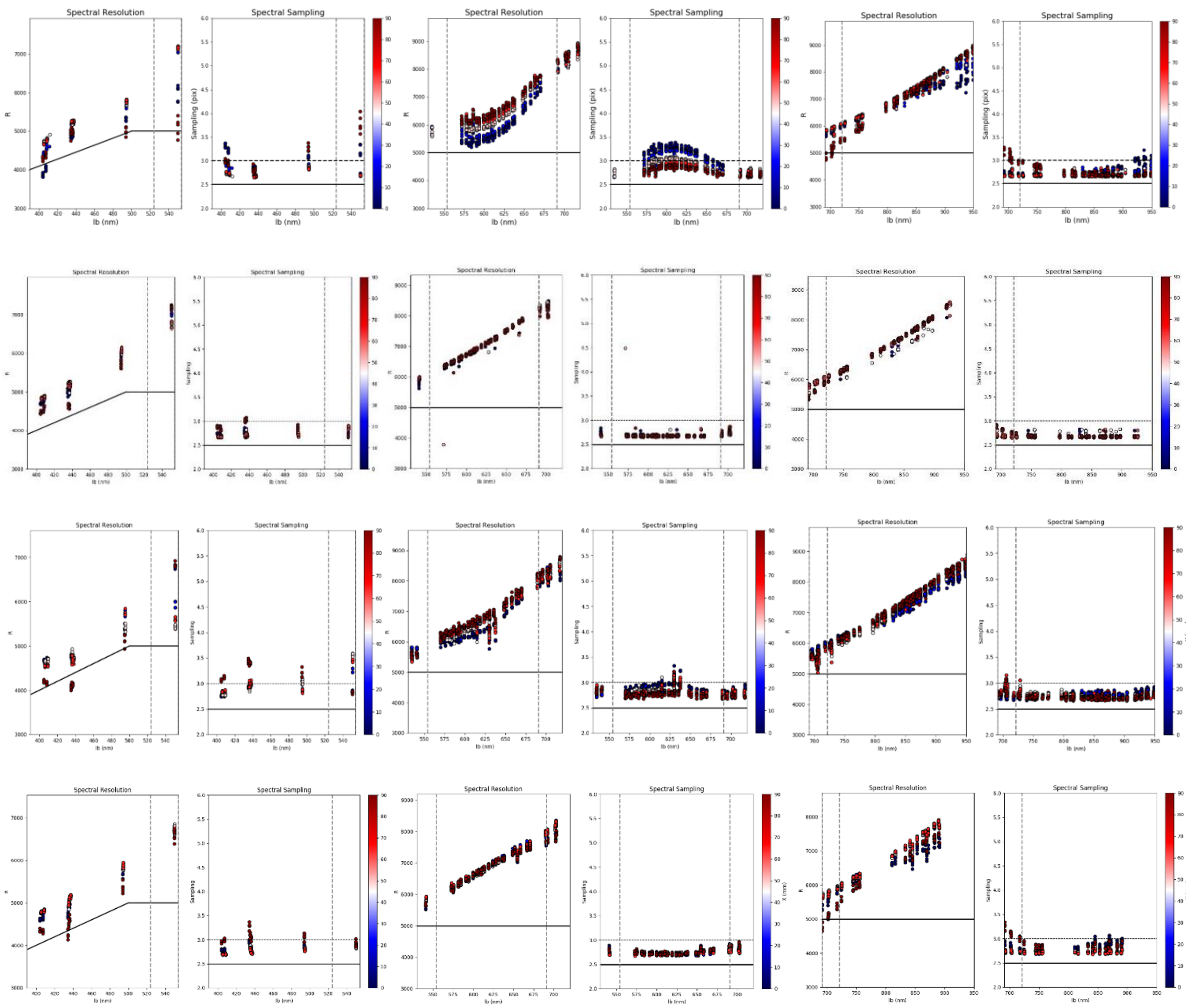

Figure 6: Spectral Resolution for LRS-A. From left to right: for Blue channel, middle for Green channel and right for Red channel. First row for MAIT@CRAL for LAR with TSU & CRAL Halogen lamp in Dec, 21. Second row for LRS-A - AIT@AIP with TSU & AIP Halogen lamp – First integration in Dec, 22. Third row: LRS-A - AIT@AIP with ESU & AIP Halogen lamp – during the LRS realignment campaign in CW49, 2024. Fourth row for LRS-A - AIV@Paranal with TSU & Other AIP Halogen lamp in July, 2025.

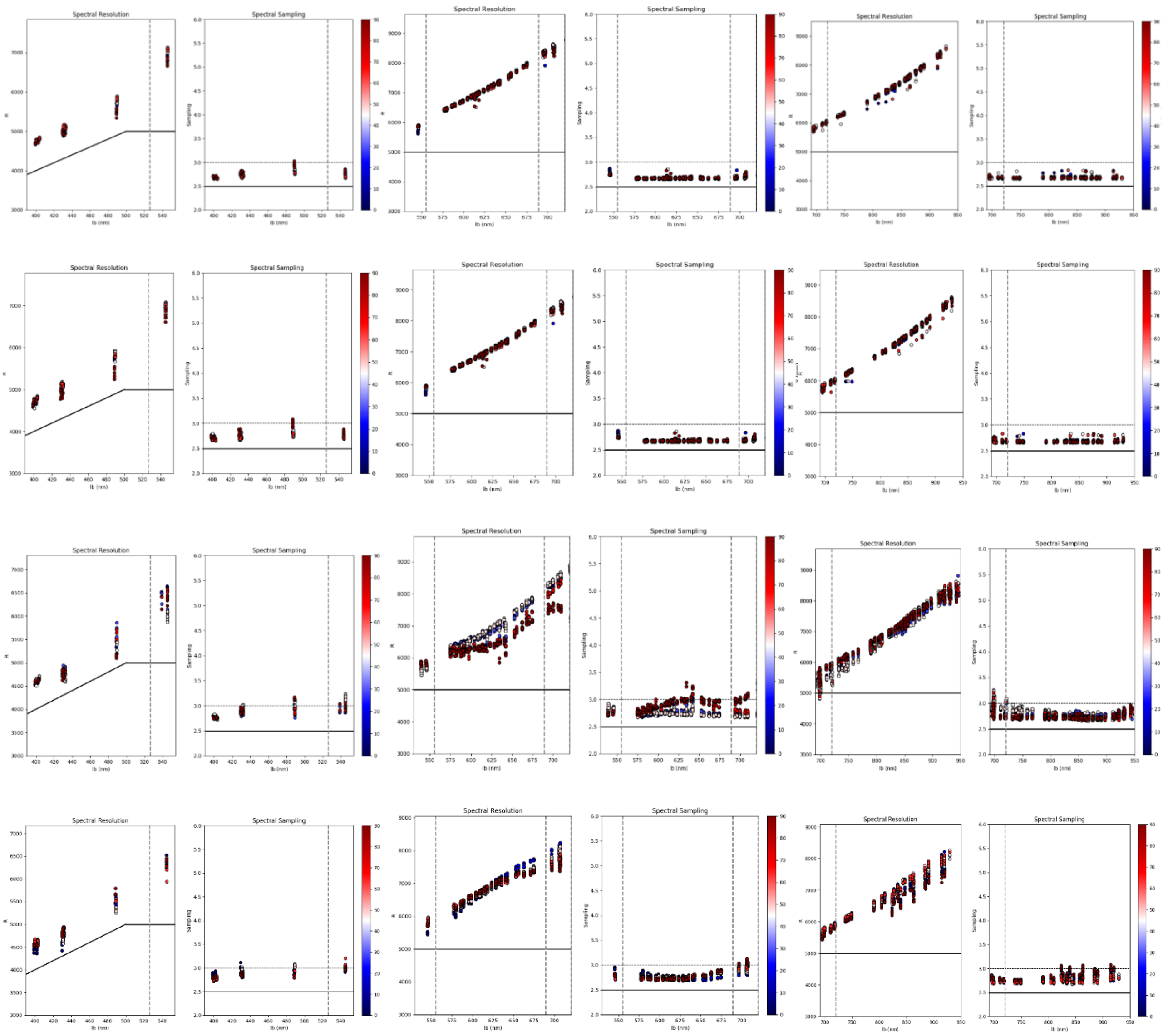

Figure 7: Spectral Resolution for LRS-B. From left to right: for Blue channel, middle for Green channel and right for Red channel. First row for MAIT@CRAL for LAR with TSU & CRAL Halogen lamp in Oct, 22. Second row for LRS-A - AIT@AIP with TSU & AIP Halogen lamp – First integration in Dec, 22. Third row: LRS-A - AIT@AIP with ESU & AIP Halogen lamp – during the LRS realignment campaign in CW49, 2024. Fourth row for LRS-A - AIV@Paranal with TSU & Other AIP Halogen lamp in July, 2025.

## 6.2 Performance using continuum lamps for the performance “Cross-talk between spectra”

A Tungsten Halogen Light Source from OceanOptics (HL-2000) was used for fibre flat exposures. The analysed exposures to produce the results contains a set of BIAS and the reference exposures. The cross-talk criterion is derived by computing the ratio between the flux contained in a 3x3 pixels box centered on the Point Spread Function (PSF), and the flux contained in the same box separated by 6.5 pixels along the spatial direction. These values are shown in Figure 8, while the mean values obtained in each arm were the following: LRS-A (Blue 0.74% , Green 0.46 %, Red 0.64 %); LRS-B (Blue 0.87%, Green 0.59 %, Red 0.60 %). The requirement, for which no spectrum was to contaminate its neighbouring spectrum on the detector with more than 2% (goal 1%) of its own flux, was fulfilled at 97.4% of the detector area in the Blue, 98.3% in the Green and 100% in the Red arm.

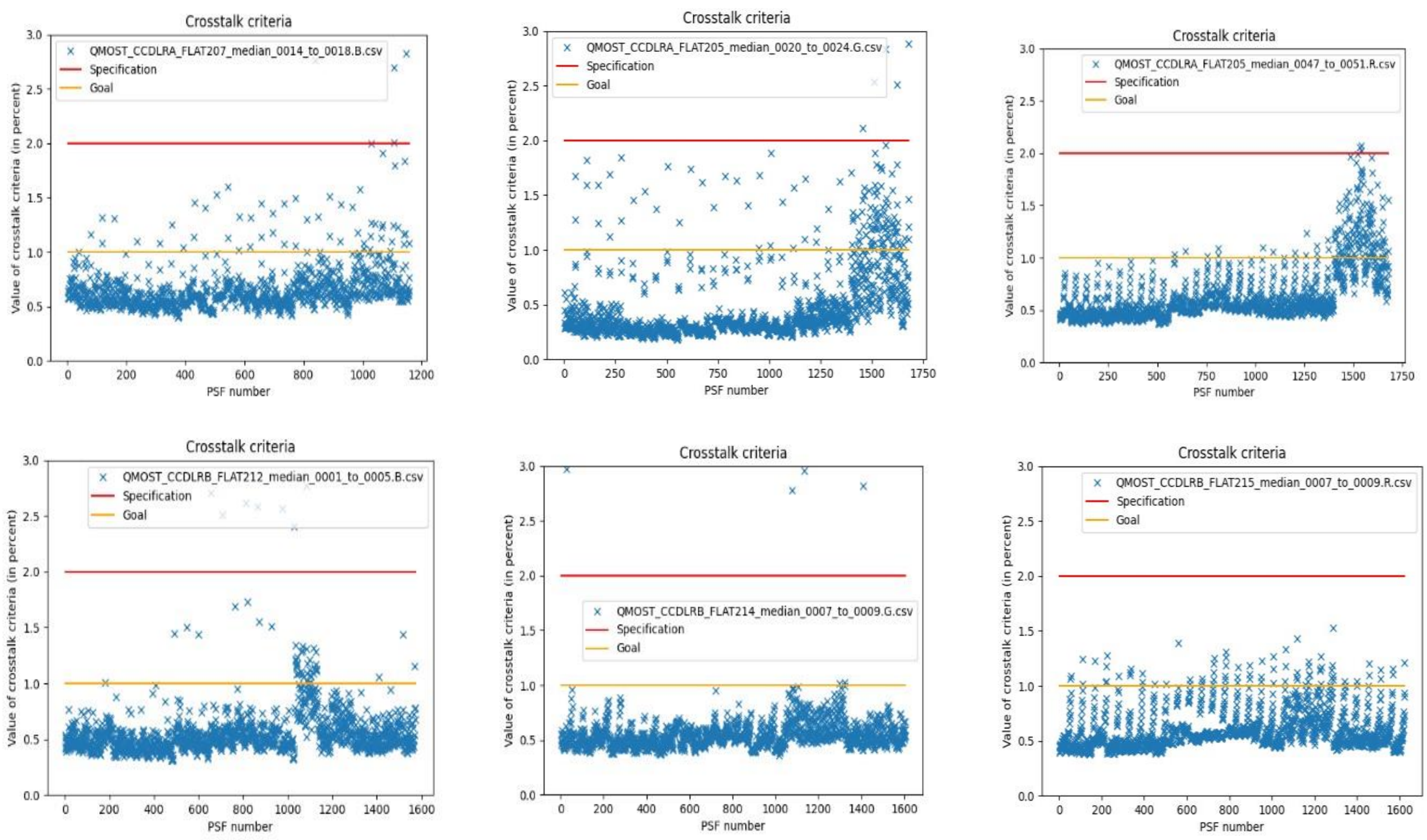

Figure 8: Cross-talk between spectra for each channel (left: blue, middle: green, right: red). Top for LRS-A and bottom for LRS-B.

In addition, the LRS-A exposures were analyzed using the pipeline developed by the Cambridge Astronomical Survey Unit (CASU) (Figure 9). This analysis revealed an improvement in alignment, confirming compliance with the specified requirements.

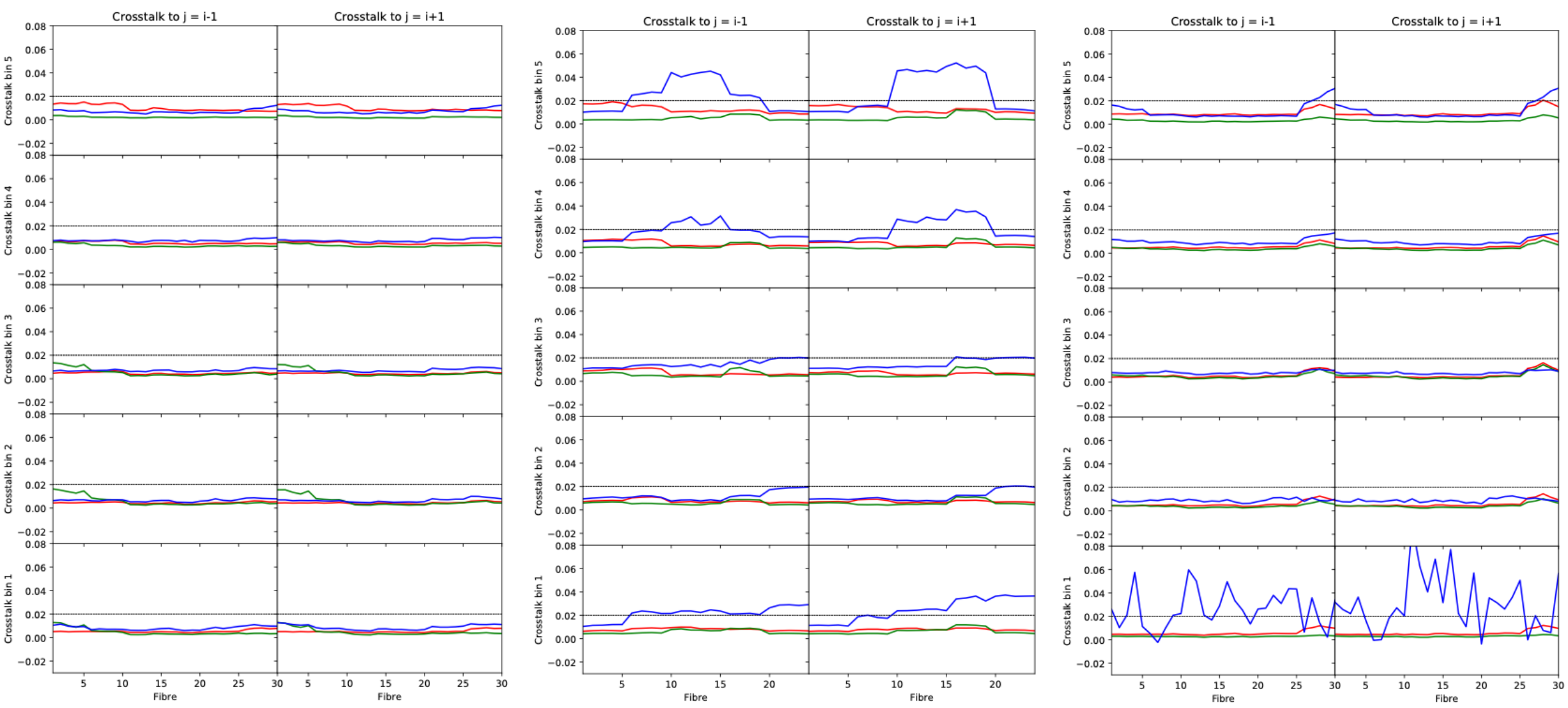

Figure 9: LRS-A Cross-talk - results from CASU pipeline in April, 22 at Potsdam (Left), December, 24 at Potsdam (Middle) and July, 25 at Paranal NIH (Right).

### 6.3 Best focus optimization

The best focus was initially based on image quality along the spatial and spectral axes, whereas the final focus position was fine-tuned based on the margins with respect to the high-level requirements which are the crosstalk values (along the spatial axis) and the spectral resolution (along the spectral axis). This resulted in the LRS-A channels to be set at best focus with a small correction (Blue = Best focus + 25µm; Green =Best focus + 25µm; Red = Best focus), while the LRS-B channels remained at the best focus position (Blue = Best focus; Green = Best focus; Red = Best focus). The crosstalk results for LRS-A and LRS-B at Paranal, compared with previous AIT phases, are shown in Figure 10, illustrating the achieved trade-off and the improvement in meeting the requirements.

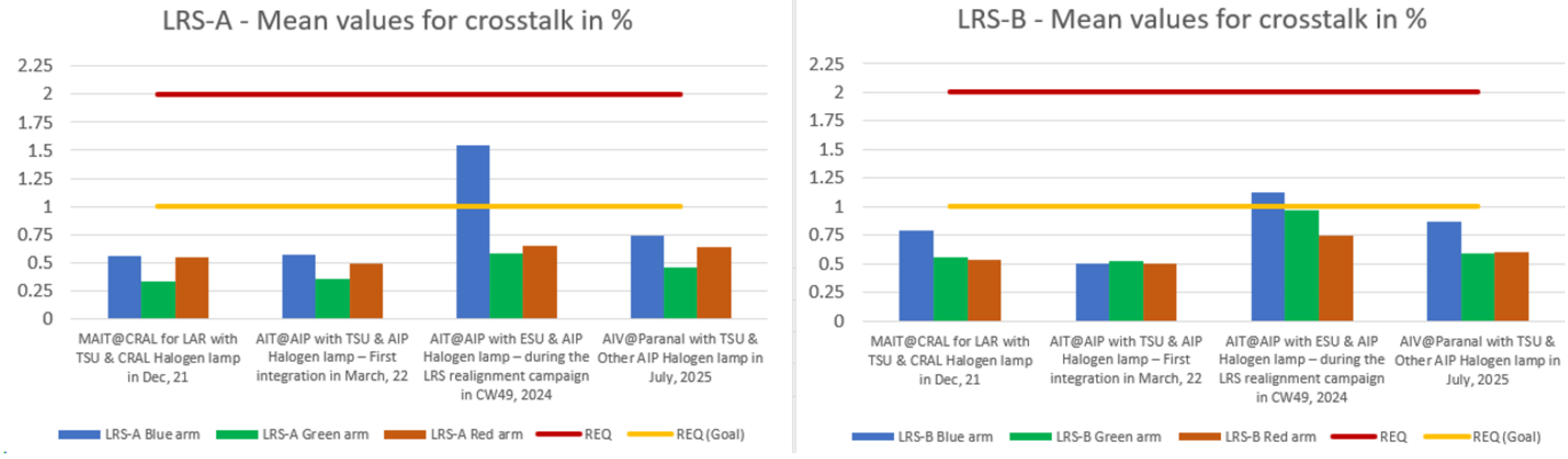

Figure 10: Mean values for Cross-talk for LRS-A and -B at different AIT phases.

## 7 LRS AIV AT VISTA TELESCOPE

The SPIE paper [9] details the process of the spectrograph installation. The author provides details the planning strategies, documentation, and stepwise integration process that ensured on-time completion, as well as the use of custom digital tools like Jira for tracking progress and test reporting. The paper concludes with lessons learned and recommendations to optimize AIV for future large-scale astronomical instruments.

### 7.1 LRS Installation on VISTA telescope

The installation of the LRSs took place in several successive steps, one day apart, following the same procedures. The LRS was picked up from the NIH and transported to the VISTA level on a truck with special shock-absorbing systems (Figure 11 and Figure 12). The LRS spent the night next to VISTA, with a truck serving as a windbreak. The installation took place early in the morning, before sunrise, to avoid glare and take advantage of calmer wind conditions.

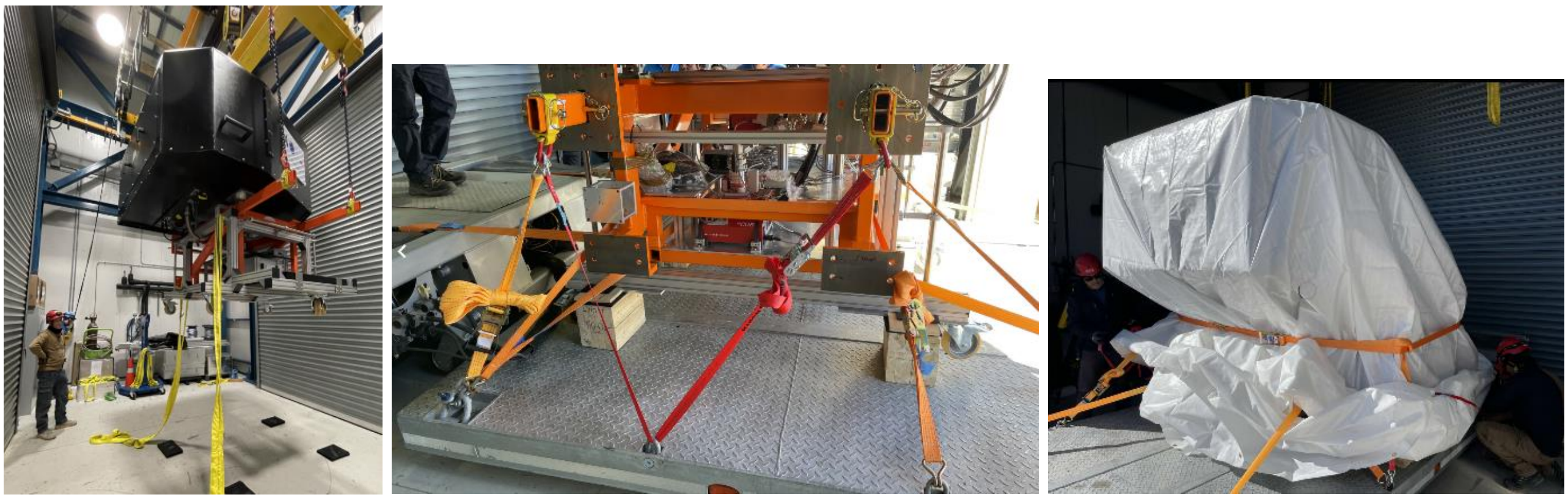

Figure 11: LRS-A at NIH ready for the transport to VISTA telescope

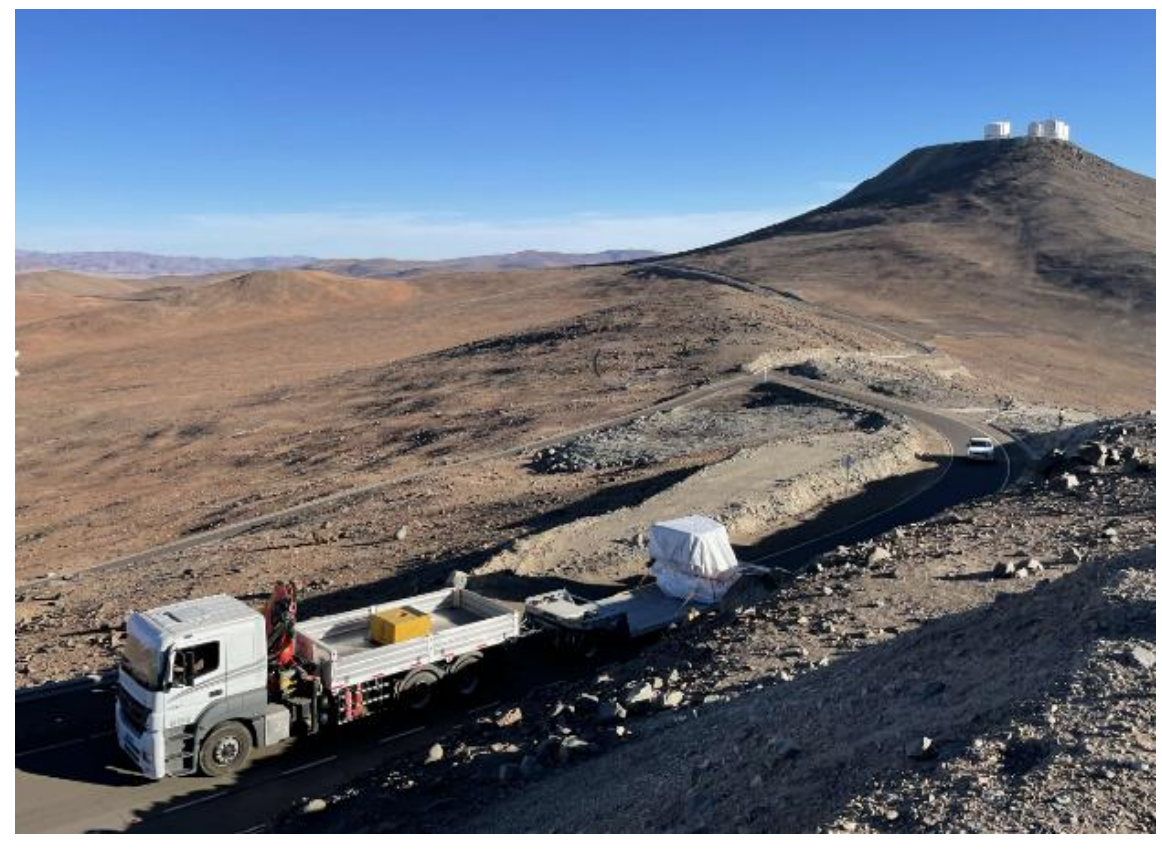
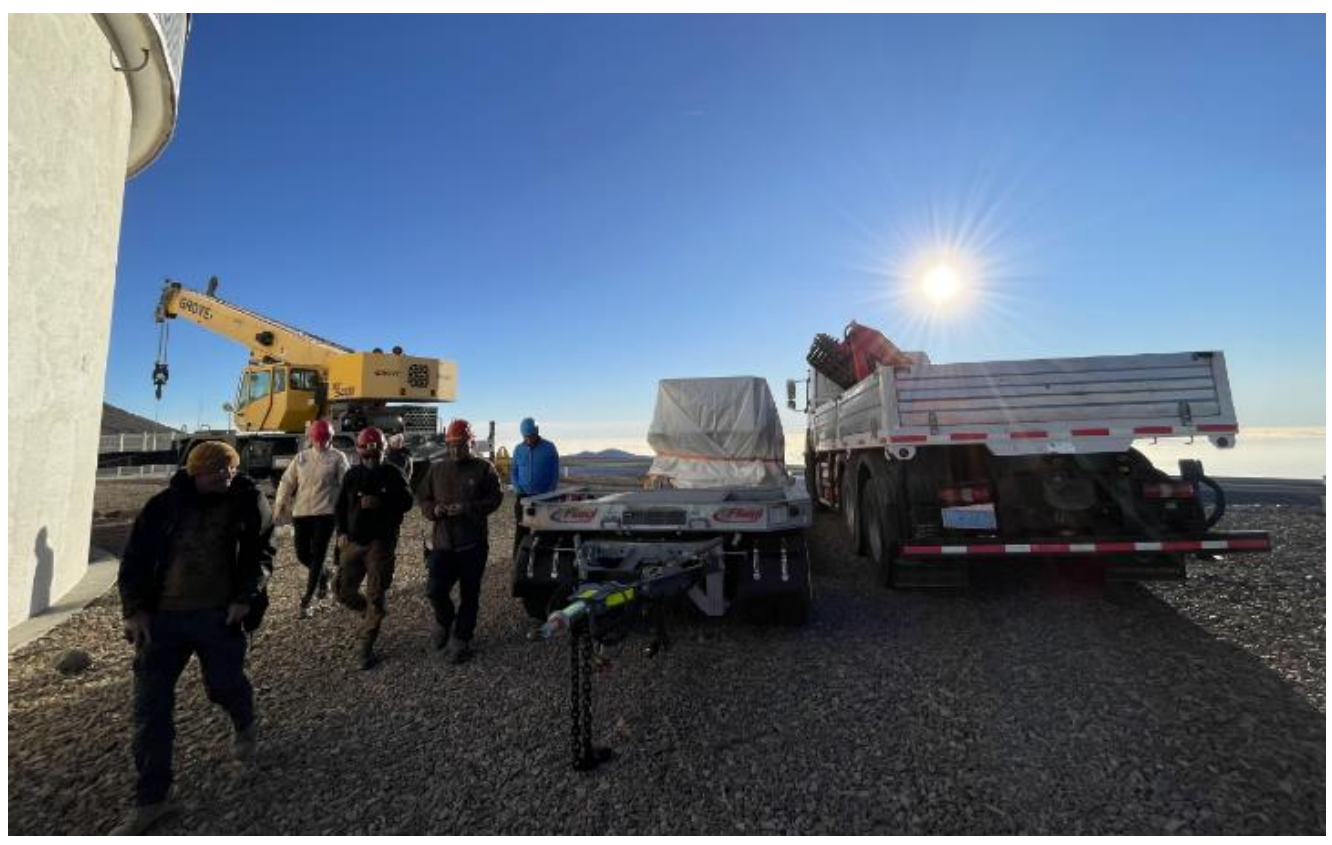

Figure 12: LRS-A transport next to VISTA telescope

At the sunrise, the LRS was picked up by the mobile crane and lifted into the VISTA enclosure. The LRS-A was the first to be test fitted on the Spectrograph Installation Platform where the interface flanges and bolts were checked. It was then pushed back and the interface flanges on the yoke were drilled for pin installation. The LRS-A was finally bolted on the telescope and the aluminum transport frame was detached (Figure 13 and Figure 14).

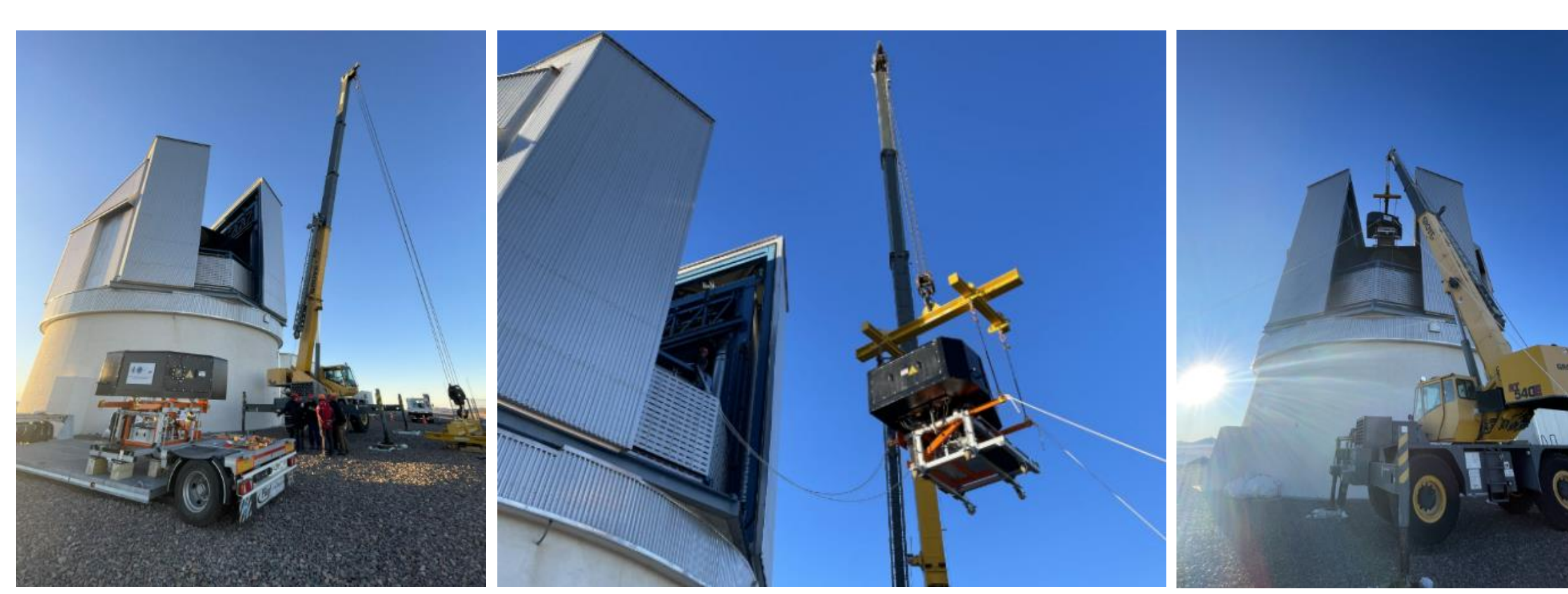

Figure 13: LRS-A outside the VISTA telescope

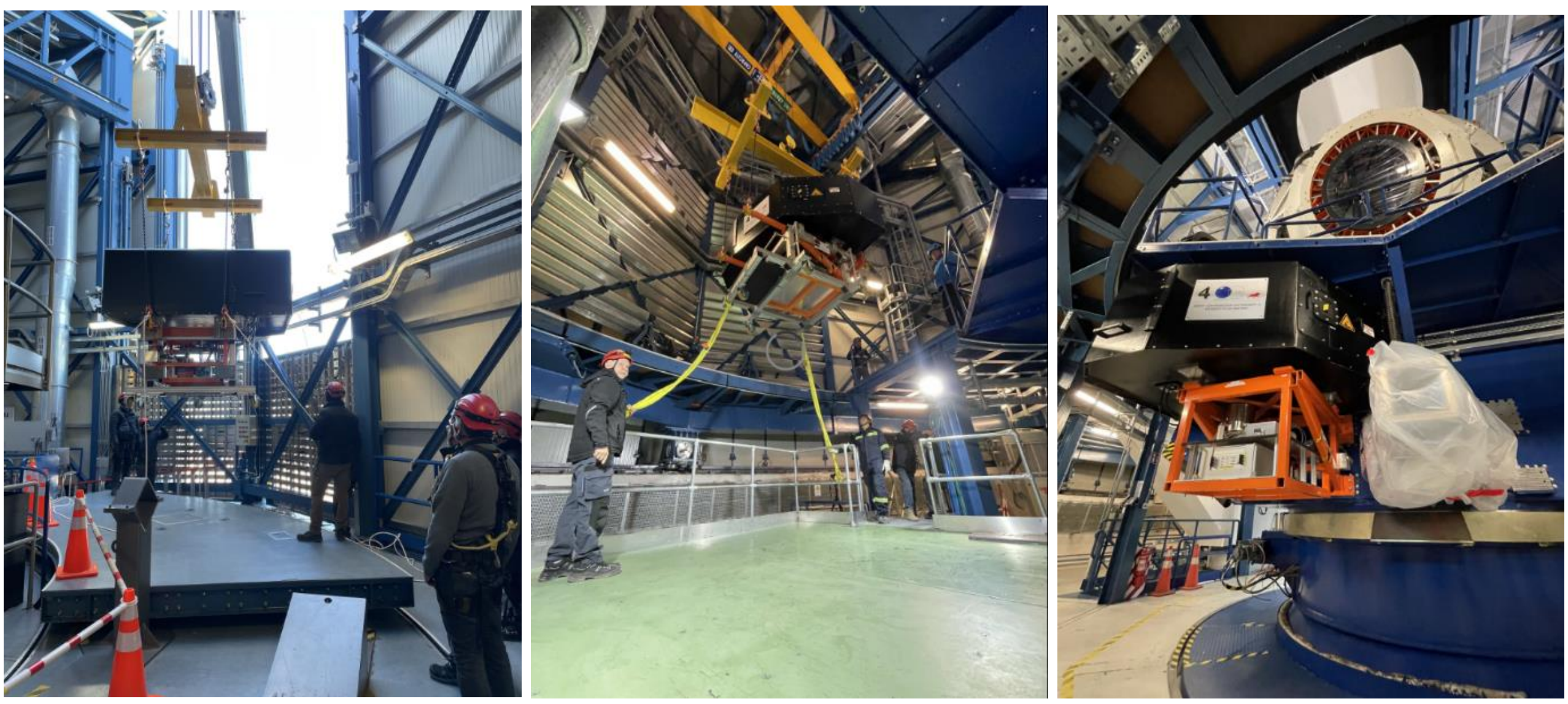
Figure 14: LRS-A inside the VISTA telescope

The LRS-B installation followed the same process. The interface matched well and the flanges were pinned. A full azimuth rotation test was performed to verify that no collisions were affecting the spectrographs (Figure 15).

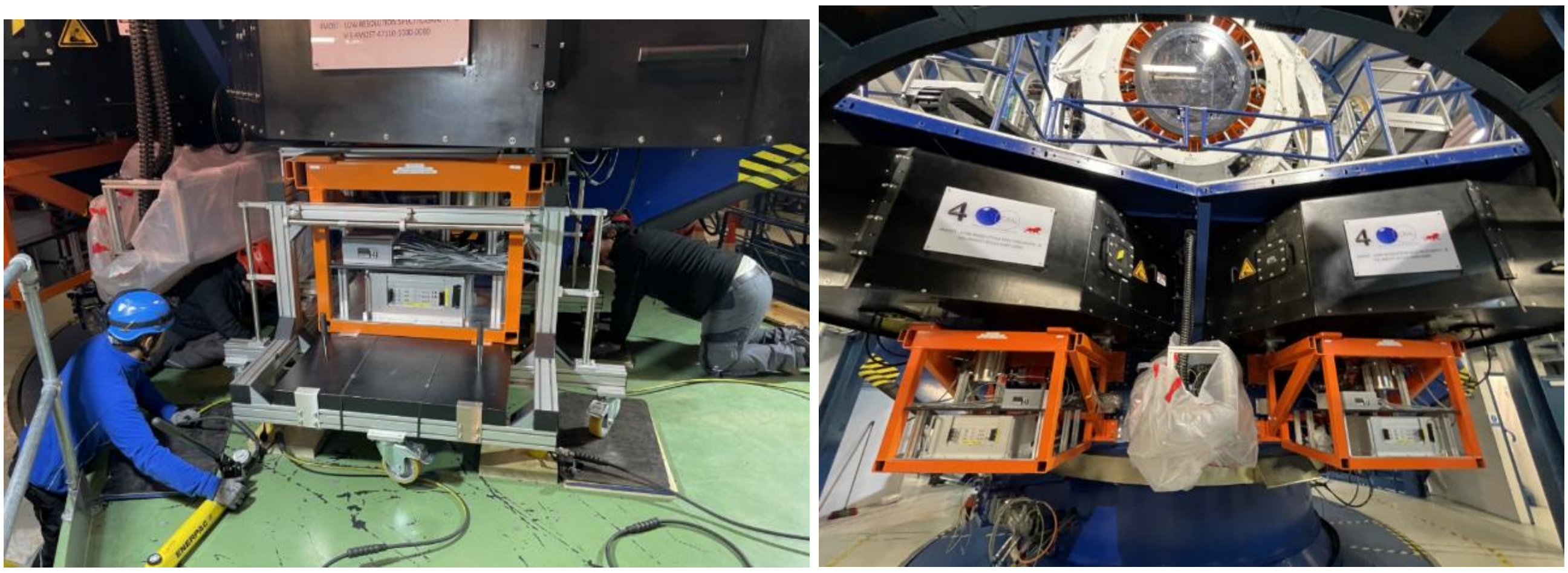
Figure 15: LRS-A and LRS-B on VISTA

## 7.2 ESU Installation in LRS and technical commissioning

The Entrance Slit Unit (ESU) were installed in the LRS-A and LRS-B, which allowed continuous fibre run from the focal surface to the spectrographs. The connections of the back-illumination and simultaneous calibration fibres were connected just after (Figure 16, Left). The Slit Temporary Frame was removed from the yoke, all electrical and fibre cables were moved aside to give good clearance to the LN2 installation.

The LRS LN2 lines were installed, connected to the VISTA's yoke and fed through inside the LRS detector volume (Figure 16, Right).

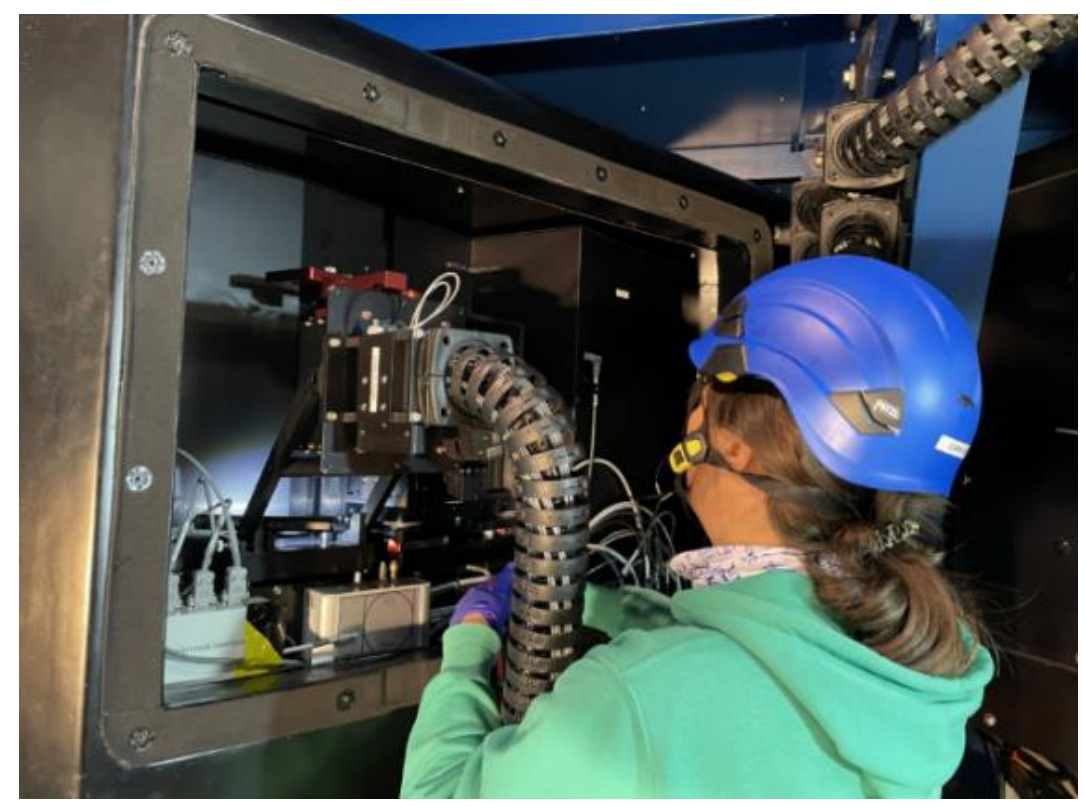

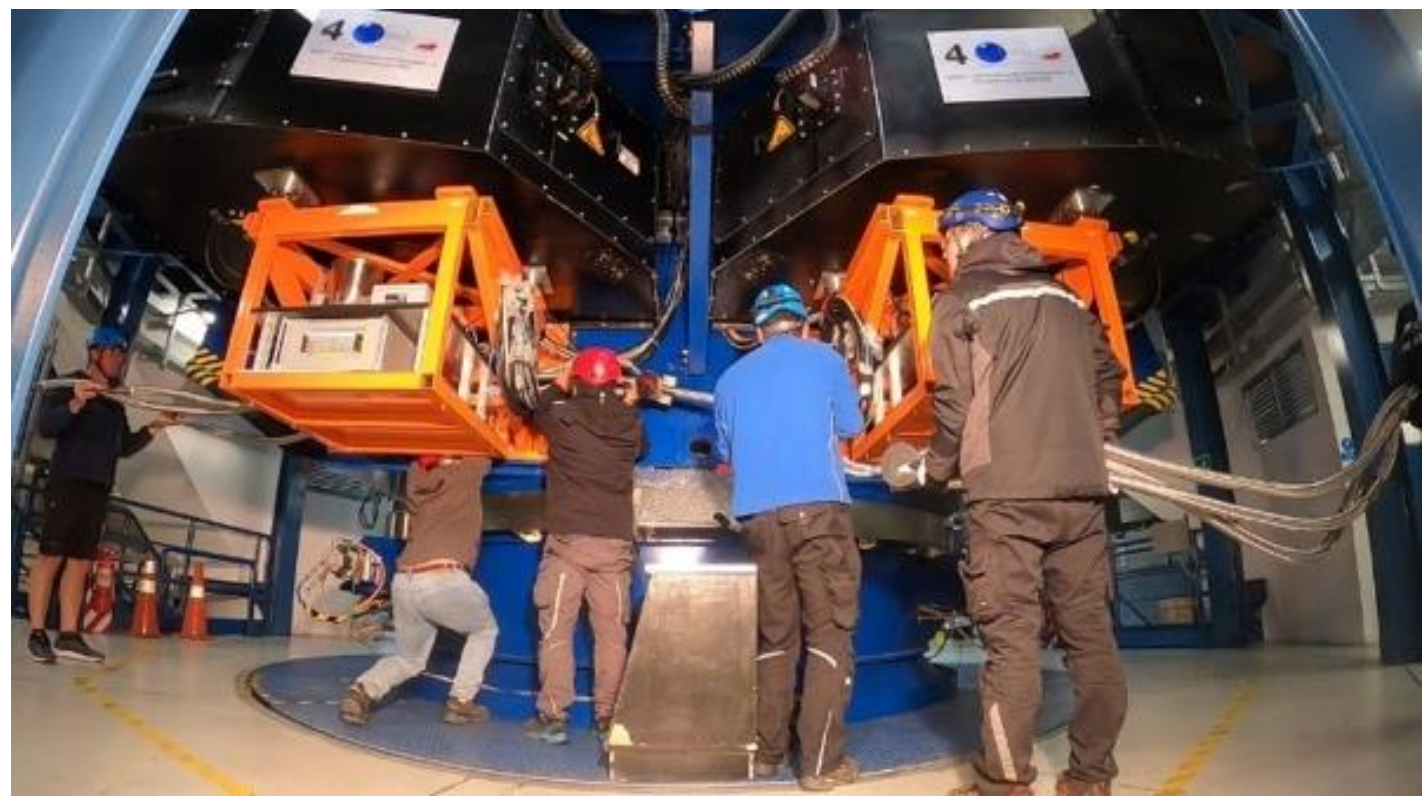

Figure 16: Left: ESU Installation in LRS. Right: LRS LN2 lines connected to the VISTA's yoke and fed through inside the LRS detector volume.

After the LN2 lines were connected to the LRSs, without dismounting the DV cover, the LRS LN2 distribution on the yoke was aligned and fixed (Figure 17, Left). The MPIA team started the cabling. Once the cabling of both LRS were completed, the LRSs were cooled down at a controlled temperature of 163 Kelvins (Figure 16, middle). After the installation of the LLCU at VISTA, the systems were analyzed in parallel to be fully integrated into the Telescope Control Network (TCN) and administered by the ESO and consortium personnel. Some initial test exposures were taken with the LRS red detectors using the dome lights. The LRS DV covers were then installed to enable the exposures verification tests (Figure 17, Right).

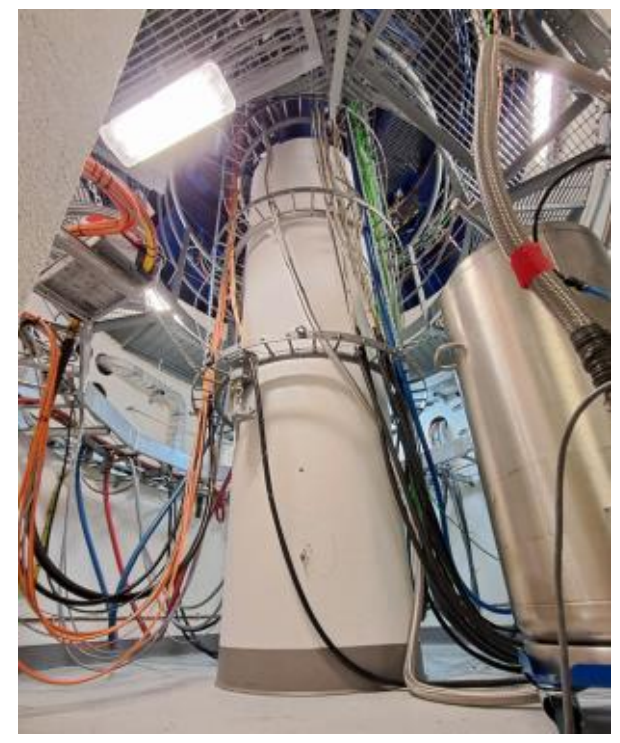

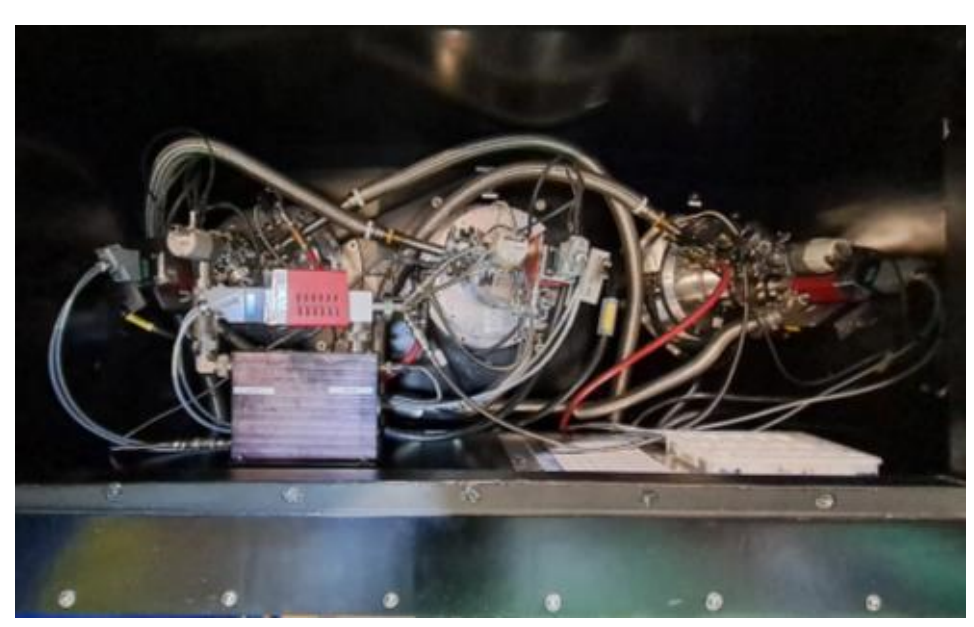

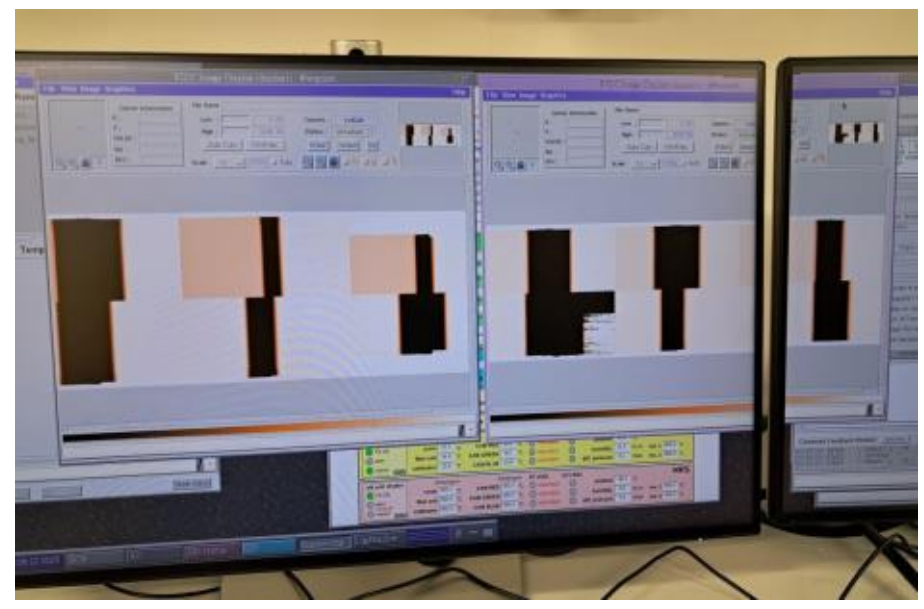

Figure 17: Left: Aligned and fixed LRS LN2 distribution on the yoke. Middle: Cabling of the LRS-A. Right: Parallel testing for integration.

### 7.3 System AIV with first "technical" exposures

Rigorous planning and preparation were crucial for the complex AIV phases of the 4MOST facility. Over ten major subsystems were successfully installed on the ESO's VISTA telescope in October 2025 with zero schedule delays. Lessons learned and recommendations to further optimize AIV for future large-scale astronomical instruments are covered in the SPIE paper [9].

The "technical" exposures were carried out to verify the proper functioning of the instrument during the technical commissioning phase. These included darks, bias, detector flats, flats with Laser Driven Light Sources (LDLS) and Fabry-Pérot, twilight, and sky exposures. Each served to check specific aspects such as detector stability, illumination uniformity, and optical alignment. These tests were not meant for scientific analysis. Special emphasis was dedicated to on-sky performance in terms of spectral resolution, cross-talk, sensitivity, and our efforts to characterize the instrument focus after installation on the telescope, which are covered in the SPIE paper [10]. The final calibration exposures, along with their detailed analysis, were carried out during the official commissioning phase.

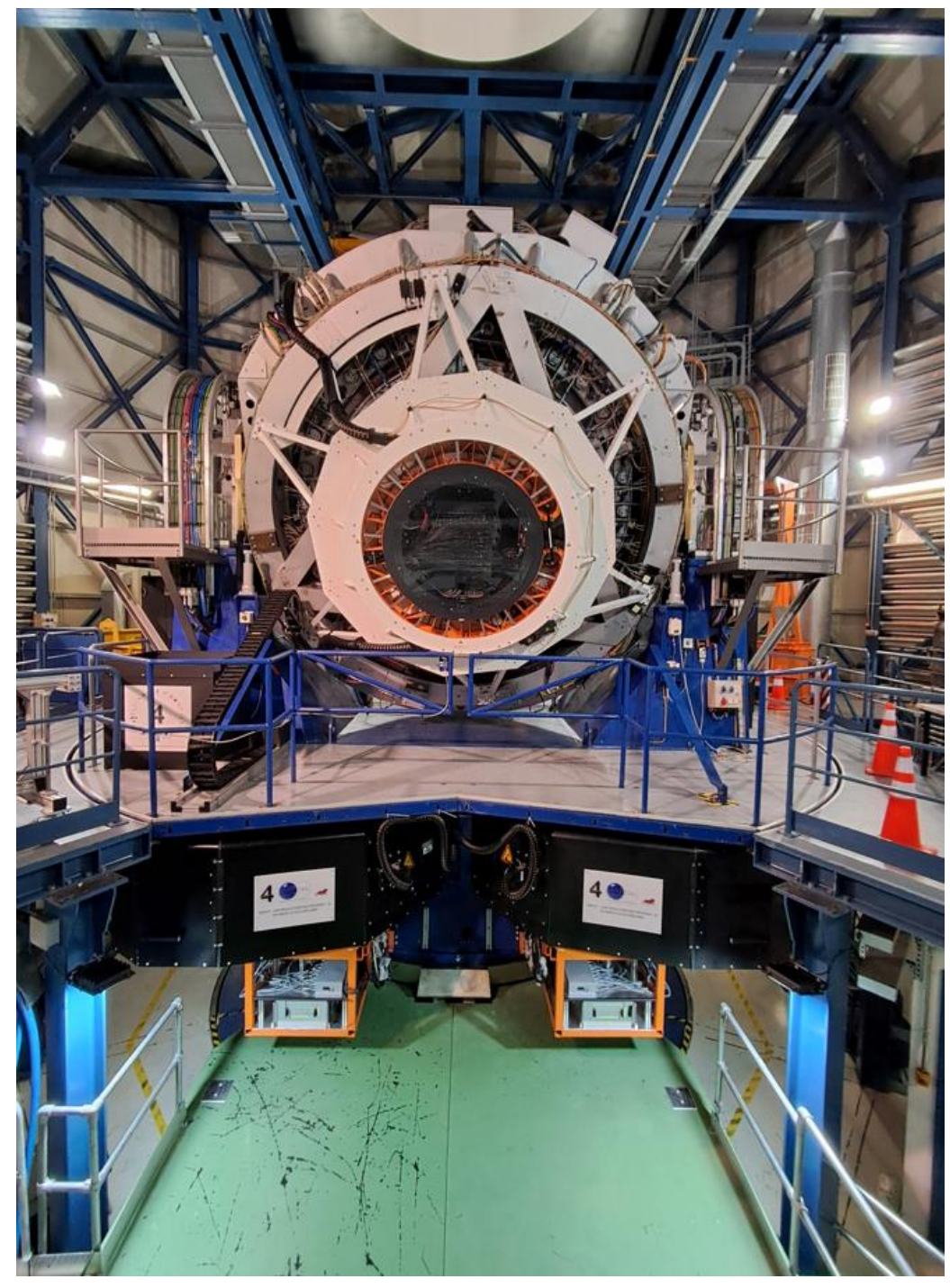

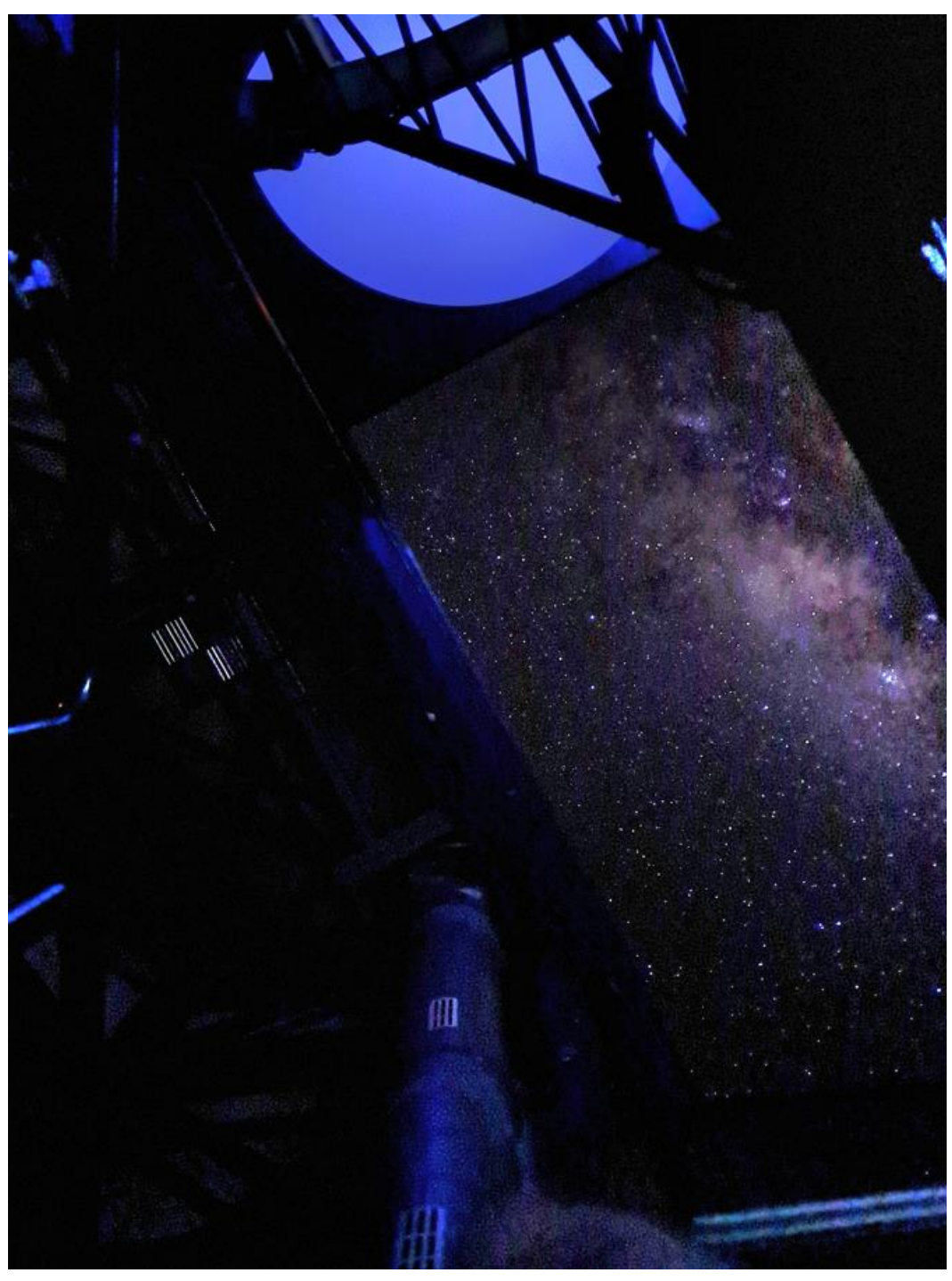

Figure 18: Left: The 2 LRSs completed on VISTA telescope. Right: VISTA telescope looking the Milky way.

# 8 CONCLUSIONS AND FUTURE DEVELOPMENT

Both LRS-A and LRS-B have been successfully aligned at the NIH in Paranal. All requirements have been met. A trade-off was done between Spectral resolution and its variation versus the crosstalk value; however, all parameters remain within acceptable limits. Their integration into the VISTA telescope, as well as the installation of the ESU were successful. The initial technical commissioning exposures showed no major issues and therefore the LRS was considered ready to move on to the official science commissioning phase.

The science commissioning, data-flow operations rehearsal, and the science verification phases will be completed in 2026. After the successful completion of the science commissioning the facility will start the first 5-year survey, which consists of 25 different science programs. An overview of instrument performance and the unique operational scheme of 4MOST is covered in the SPIE paper [11].

# ACKNOWLEDGMENTS

I would like to particularly thank the members of the 4MOST CRAL team, ESO, AIP Project Office members and everyone working on the 4MOST project for their support and invaluable contribution to this encouraging 4MOST LRS AIV phase.

We acknowledge the financial support from CRAL, the Commission Spécialisée Astronomie-Astrophysique (CSAA) of CNRS/INSU, Université de Lyon, Université de Lyon1 and the LabEx Lyon Institut of Origin (LIO).

# REFERENCES

[1] Steffen Frey et al., "4MOST preliminary instrument design," Proc SPIE 9908, 310 (2016)

[2] Olga Bellido-Tirado et al., " 4MOST systems engineering: from conceptual design to preliminary design review," Proc SPIE 9911, 76 (2016)

[3] Florence Laurent et al., "4MOST Low Resolution Spectrograph MAIT," Proc SPIE 11447/12.2561487 (2020)

[4] Andreas Kelz et al., " 4MOST: manufacture, assembly and test of the optical fiber system," Proc SPIE 12184-263 (2022)
[5] Florence Laurent et al., "4MOST Low Resolution Spectrograph Alignment," Proc SPIE 12184-255 (2022)
[6] Karen Disseau et al., "4MOST Low Resolution Spectrograph Performance," Proc SPIE 12184-272 (2022)
[7] Florence Laurent et al., " 4MOST Low Resolution Spectrograph AIT at AIP", Proc SPIE 13096-241 (2024)
[8] Micheva, Genoveva et al., "Preliminary characterization and verification of 4MOST in Europe," Proc SPIE 13096-75 (2024)
[9] Allar Saviauk et al., " The backbone of success: detailing the planning and execution of the 4MOST installation on VISTA telescope", Proc SPIE 14152-56 (2026)
[10] Jean-Kristian Krogager et al., "4MOST: on-sky performance of the Low Resolution Spectrographs (LRS)", Proc SPIE 14149-112 (2026)
[11] Jong, Roelof S et al., " 4MOST: the 4-metre Multi-Object Spectroscopic Telescope project at the start of science operations", Proc SPIE 14149-48 (2026)